# Fabrication of $GaN_xAs_{1-x}$ Quantum Structures by Focused Ion Beam Patterning

K. Alberi[1,2], A. Minor[3], M.A. Scarpulla[1,2], S.J. Chung[1], D.E. Mars[4], K.M. Yu[2], W. Walukiewicz[2], and O.D. Dubon[1,2]

[1]*Department of Materials Science and Engineering, University of California, Berkeley, CA 94720*
[2]*Lawrence Berkeley National Laboratory, Berkeley, CA 94720*
[3]*National Center for Electron Microscopy, Lawrence Berkeley National Laboratory, Berkeley, CA 94720*
[4]*Agilent Laboratories, Palo Alto, CA 94304*

**Abstract.** A novel approach to the fabrication of $GaN_xAs_{1-x}$ quantum dots and wires via ion beam patterning is presented. Photomodulated reflectance spectra confirm that N can be released from the As sublattice of an MBE-grown $GaN_xAs_{1-x}$ film by amorphization through ion implantation followed by regrowth upon rapid thermal annealing (RTA). Amorphization may be achieved with a focused ion beam (FIB), which is used to implant Ga ions in patterned lines such that annealing produces GaAs regions within a $GaN_xAs_{1-x}$ film. The profiles of these amorphized lines are dependent upon the dose implanted, and the film reaches a damage threshold during RTA due to excess Ga. By altering the FIB implantation pattern, quantum dots or wires may be fabricated.

## INTRODUCTION

The $GaN_xAs_{1-x}$ system belongs to a class of materials known as highly mismatched semiconductor alloys (HMA) in which a small amount of the anion is replaced by an isovalent impurity of a much different electronegativity. Anticrossing effects between the localized impurity states and the delocalized states of the matrix result in the lowering of the conduction band edge, as described by the band anticrossing model [1]. In the case of $GaN_xAs_{1-x}$, substitution of 1 mol % of N on the As sublattice results in a reduction of the bandgap by as much as 180 meV [2]. The ability to manipulate the bandgap by carefully controlling small concentrations of N makes this system well-suited for the formation of quantum wires and dots by alternative synthesis routes. Here we introduce a technique to construct quantum structures from a $GaN_{0.015}As_{0.985}$ film using a focused ion beam.

## EXPERIMENTAL

A focused ion beam (FIB) operated at an accelerating voltage of 30 keV and a beam current of 1pA was used to selectively amorphize patterned lines by implanting $Ga^+$ ions at a 7° angle to the normal into a 10nm thick $GaN_{0.012}As_{0.988}$ film grown by molecular beam epitaxy on a semi-insulating GaAs substrate. Upon rapid thermal annealing (RTA), N is released from the As sublattice as these amorphous regions undergo solid-phase epitaxial regrowth, producing GaAs barriers surrounding the unimplanted $GaN_{0.012}As_{0.988}$ regions. Depending on the implanted pattern, quantum wires or dots are generated.

The release of N from the lattice was studied by measuring with photomodulated reflectance (PR) spectroscopy the change in the bandgap at room temperature in $Ar^+$ implanted $GaN_{0.012}As_{0.988}$ films both before and after implantation and an RTA cycle at 800 °C for 15s. The $Ar^+$ implant was carried out at room temperature with an accelerating voltage of 32 keV and a dose of $1x10^{15}$ ions/cm$^2$ and was used to simulate the amorphization conditions achieved by FIB implantation.

Cross-sectional transmission electron microscopy (XTEM) analysis of the implant profile was carried out on FIB-patterned GaAs substrates and

$GaN_{0.012}As_{0.988}$ films both before and after an RTA at 800 °C for 15s. The test structure consisted of 9 lines of implant doses ranging from $10^{13}$ to $10^{15}$ ions/cm$^2$, which were controlled by varying the ion beam scan times. The XTEM specimens were prepared in the FIB by a lift-out technique, and the surface of the sample was protected with a Pt layer deposited with the electron beam prior to ion milling.

## RESULTS AND DISCUSSION

PR measurements of the $GaN_xAs_{1-x}$ film prior to Ar+ ion implantation yield two transitions, one due to the GaAs substrate at 1.42 eV and another due to the film at 1.20 eV, indicating that the N content was around 1.2 mol %. PR measurements taken directly after Ar+ implantation show an absence of these transitions, while subsequent measurements taken after RTA reveal only the pure GaAs bandgap transition at 1.42 eV. These results indicate that it is possible to release N from the lattice of an MBE grown $GaN_xAs_{1-x}$ film through amorphization by ion implantation followed by rapid thermal annealing.

Fig. 1 displays TEM micrographs of two amorphized lines in a GaAs substrate, used for preliminary investigation, with implant doses of $3 \times 10^{13}$ and $2 \times 10^{14}$ ions/cm$^2$, respectively. The dark regions near the surface reveal the FIB-induced damage caused by the implantation of Ga ions. The spatial extent of amorphization depends on the implant dose, which is presumably due to the Gaussian distribution of the ion beam flux.

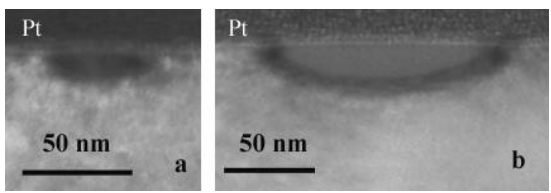

**FIGURE 1.** Cross-sectional TEM images of a) Line amorphized with a Ga$^+$ implant dose of $3 \times 10^{13}$ ions/cm$^2$, b) line amorphized with a Ga$^+$ implant dose of $2 \times 10^{14}$ ions/cm$^2$

Scanning electron microscopy of regrown lines in a $GaN_{0.012}As_{0.988}$ film, which were implanted with the Ga doses mentioned above and subsequently RTA'd, reveals that there is a surface damage threshold reached during RTA above a certain implant dose. The secondary electron image in Fig. 2, taken in the FIB after RTA, clearly shows that pitting begins to form in and around the implanted regions after RTA for implant doses exceeding $6 \times 10^{13}$ ions/cm$^2$. This damage can be eliminated by etching the surface with HCl prior to RTA to remove excess Ga, which acts to dissolve GaAs at elevated temperatures. Preliminary XTEM analysis indicates that regions amorphized with implanted Ga+ doses up to $3 \times 10^{13}$ ions/cm$^2$ do regrow under the RTA conditions described above.

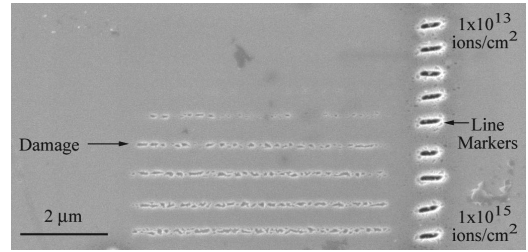

**FIGURE 2.** Secondary electron image of damage caused by RTA. The doses of the implanted lines range from $1 \times 10^{13}$ ions/cm$^2$ to $1 \times 10^{15}$ ions/cm$^2$, with damage becoming visible around $6 \times 10^{13}$ ions/cm$^2$.

## CONCLUSION

The results presented herein establish a novel fabrication method for producing quantum wires and dots. First, we have shown that N can be removed from a $GaN_xAs_{1-x}$ film via amorphization and regrowth by rapid thermal annealing. Ion implantation using a FIB followed by RTA was then introduced as a viable method for amorphizing and regrowing selected regions of a film in such a way that quantum structures consisting of GaAs barriers surrounding regions of $GaN_xAs_{1-x}$ can be constructed. Future investigation will focus on manipulating the shape of the barriers and studying their effects on the formation and performance of quantum structures.

## ACKNOWLEDGMENTS

This work was supported in part by the Director, Office of Science, Office of Basic Energy Sciences, Division of Materials Sciences and Engineering, of the U.S. Department of Energy under Contract No. DE-AC03-76SF00098. This work was supported in part by NREL under subcontract No. XaX-3-33647-01.